# A CRISPR-Cas-Inspired Mechanism for Detecting Hardware Trojans in FPGA Devices

**Dillon Staub[1], Rashmi Jha[1], and David Kapp[2]**
[1]Department of Electrical Engineering and Computer Science, University of Cincinnati, Cincinnati, OH 45220 USA
[2]Air Force Research Laboratory, Wright-Patterson AFB, OH, 45433 USA

This work was supported in part by the Air Force Research Laboratory under AFRL Award # FA8650-18-C-1191.

**ABSTRACT**

Hardware security has risen in prominence in recent years with concerns stemming from a globalizing semiconductor supply chain and increased third-party IP (intellectual property) usage. Trojan detection is of paramount importance for ensuring systems with confidentiality, integrity, and availability. Existing methods for hardware Trojan detection in FPGA (field programmable gate array) devices include test-time methods, pre-implementation methods, and run-time methods. The first two methods provide effective ways of detecting some Trojans; however, Trojans may be specifically designed to avoid detection at test-time or before implementation making run-time detection a more attractive option. Run-time detection and removal of Trojans is highly desirable due to the wide range of critical systems which are deployed on FPGAs and may be difficult or costly to remove from operation. Many parallels can be drawn between hardware and natural systems, and one example creates an analogy between hardware attacks and biological attacks. We propose a CRISPR-Cas-inspired (clustered regularly interspaced palindromic repeats) method for detecting hardware Trojans in FPGAs. The fundamental concepts of the Type 1-E CRISPR-Cas mechanism are discussed and simulated to predict the flow of genetic information through this biological system. The basic structure of this system is utilized to propose a novel run-time Trojan detection method titled CADEFT (CRISPR-Cas-based Algorithm for DEtection of FPGA Trojans). Different levels of FPGA application design flow are explored, and CADEFT is proposed for realization at the bitstream level to monitor the configuration bitstream and the run-time properties of the FPGA. The flexibility of CADEFT originates in the CRISPR-Cas mechanism's ability to recognize similar albeit previously unseen patterns which may pose a threat to the system.

## I. INTRODUCTION

The impact of electronic systems on society and modern life cannot be overstated. At its core, this continuing technological transformation is driven by microelectronic devices. With such a profound reach into the very fabric of society and civilization, the safety, security, and reliability of such devices and systems are paramount concerns. The significance of software security research has long been apparent; however, the last two decades have seen many organizations realize the implications of vulnerabilities in the hardware domain. Accordingly, research in both hardware and software security fields has expanded tremendously with renewed interest in potential hardware vulnerabilities and mitigation mechanisms. There is a push to better understand the limitations and weaknesses of hardware systems especially in the wake of several major revelations regarding hardware vulnerabilities including Meltdown, Rowhammer, Spectre, and Spoiler, all examples of recent state-of-the-art microarchitectural attacks [1]. Threats to hardware systems are diverse and range from hardware Trojans to IP piracy and reverse engineering [2]. As such, proposed defense and protection methods span a broad range of mechanisms including split manufacturing and fingerprinting among others.

With new or modified attack and defense methods continuously being introduced, the competition between designers and attackers, which has been likened to a game of 'cat and mouse,' will persist. Especially within the realm of complex networks and systems, adaptive systems have long been the subject of much study and admiration. The ability to react and change dynamically based on internal or external stimuli such as environmental variations or a predator's attack is a foundational aspect of biological organisms' survival. This adaptivity is a coveted characteristic of biological systems which has applications not only in hardware security but also across a broad spectrum of electronic systems.

## A. BIOLOGICAL EVOLUTION & ADAPTATION

The natural world contains many systems, both simple and complex, which are valuable sources of knowledge and inspiration for fields ranging from engineering to medicine to arts. One of nature's most intriguing and heavily studied areas is evolution and the variety of mechanisms which make it possible. Evolution concerns the change in an organism's characteristics over multiple generations via a selection process. Generally, offspring with characteristics more favorable to their environment have a higher likelihood of survival and of passing their characteristics to the next generation. While knowledge of evolution may be incomplete, it has inspired myriad approaches to challenges in areas like computer security, image identification, and natural language processing among others.

Although evolution plays a central role in the adaptivity and survival of biological organisms, it is not alone. Natural evolution is a tremendous resource because billions of years of evolution have produced not only myriad organisms but also a wide array of adaptive systems within these organisms. Even within prokaryotic microorganisms, several systems exist for defense and adaptation purposes including restriction modification (RM) systems, CRISPR-Cas systems, abortive infection systems, and lateral gene transfer among others.

Certain microorganisms may possess more than one of these systems granting it multiple opportunities to adapt and survive in the face of threats. Altogether, these cellular defense and adaptation systems are powerful mechanisms honed through evolutionary processes which help to protect the organism and increase its chances of survival. Despite their relationship to one another, one major distinction can be made between evolution and these defense and adaptation systems. Evolution concerns changes which occur over multiple generations, or intergenerational changes; however, the changes induced via the defense and adaptation systems previously described normally occur during a single generation and can be classified as intragenerational changes.

## B. EVOLUTIONARY ALGORITHMS

Evolutionary algorithms refer to a broad subset of optimization algorithms which draw inspiration from biological evolution mechanisms like reproduction, mutation, and selection. In general, these algorithms rely on "survival of the fittest" criteria whereby organisms or instances are evolved via one of the aforementioned mechanisms and their fitness is calculated in some manner. This fitness metric is used to make a determination regarding the organism's survival and ability to pass on its characteristics to future generations. While CRISPR-Cas mechanisms and evolution are certainly related, there is an important distinction. Evolution and algorithms based on evolutionary characteristics demonstrate adaptation between generations; however, CRISPR-Cas systems enable adaptation within a single generation. From a biological standpoint, this is significant because it allows an organism like *Escherichia coli* to adapt to threats during its lifetime and pass these characteristics on to its offspring instead of relying solely on mechanisms like reproduction, mutation, or selection to enable the offspring to survive similar threats. CRISPR's relationship with evolution and reproduction is important as the efficacy of the mechanism is enabled in part by the ability to pass down to offspring acquired genetic information regarding threats to the cell's survival. The close natural relationship between the CRISPR-Cas mechanism and evolution indicates that a synthetic implementation of a cohesive system should be not only possible but also effective.

## C. FPGAs

Field-programmable gate arrays, or FPGAs, are designed as flexible hardware platforms which can be configured for a wide variety of applications. This versatility along with growth in FPGA capacity and complexity has led to their adoption in dozens of fields including aerospace, automotive, bioinformatics, consumer electronics, high-performance computing, medical, and wireless communications among others [3].

With their widespread usage across industries, valid concerns have been raised regarding the security of these systems. The versatility of FPGAs is a drawback from this perspective as hardware vulnerabilities which affect one FPGA are likely effective across many or all other FPGA devices. The distribution of these systems means that a Trojan or other hardware attack could potentially disrupt critical systems or infrastructure resulting in a catastrophe for companies or even countries. In efforts to mitigate these risks, research into hardware security and FPGA Trojans and vulnerabilities specifically has increased significantly, and a wide variety of detection and tolerance strategies have been proposed to mitigate or eliminate both Trojans and the issues they cause.

## D. HARDWARE SECURITY

Hardware security is a broad term referring to the protection of physical devices from vulnerabilities. These vulnerabilities can manifest as IP piracy, counterfeiting, hardware Trojans, physical tampering, side-channel attacks, and fault attacks among others. Methods for ensuring hardware security are varied, and implementation options and details depend greatly on the electronic system being studied; however, at its core, hardware security concerns security and trust from a hardware perspective. Hardware security is a hot topic in the electronic systems industry due to its relatively recent emergence into the collective psyche of the industry. Throughout hardware supply chains, vulnerabilities are being sought out and investigated to prevent potential hardware attacks and other issues like overproduction and counterfeiting.

## E. CONNECTING BIOLOGY TO HARDWARE SECURITY

When comparing nature to artificial systems, many parallels can be drawn. One example of this lies in the relationship between microorganisms and bacteriophages and the relationship between hardware systems and hardware attacks. A quintessential "cat-and-mouse game", this relationship is a story of co-evolution and adaptation as one organism develops a mechanism or mechanisms which outmaneuvers the mechanisms of the other organism. In the case of a microbe, this could be the development of an immune-like response to the detection of a specific bacteriophage. In the case of a bacteriophage, it could be the modification of specific proteins to prevent binding or interference by the microbe's systems.

Biologically-inspired defense mechanisms have been studied in detail for use in cyber and software systems; however, their utility in hardware systems has yet to attract comparable attention. Several works including [4,5,6] detail potential hardware defense systems based on natural phenomena, but to the author's knowledge, no work has focused on developing a hardware defense mechanism based on CRISPR-Cas as a sub-cell process.

### F. THIS PAPER

This paper begins by examining the biological mechanisms of the CRISPR-Cas adaptive immunity system. It summarizes the phases of the Type 1-E CRISPR-Cas system present in *Escherichia coli* and significant components of this system. Next, a biological model is presented which explains the flow of genetic information through the CRISPR-Cas system, and simulations are presented to demonstrate the genetic mechanisms which enable the system to effectively regulate foreign genetic material. Hardware Trojans are discussed in greater detail including methods used to prevent, detect, and remove them from FPGA systems. The FPGA design flow is described and examined for the feasibility of Trojan insertion and realization of schemes for Trojan detection. In connecting biology to hardware security in an explicit manner, CRISPR-Cas-inspired methods for detection of Trojans in FPGA devices are proposed and described. Finally, this paper concludes with a discussion of future work in FPGA Trojan detection.

## II. CRISPR-CAS INTRODUCTION

Over the past two decades, the biology world has been abuzz with the implications of Clustered Regularly Interspaced Palindromic Repeats, or CRISPR, systems for use in both research and industry. Recent years have seen significant investment and research interest in developing functional, controllable CRISPR systems. With the potential to edit genetic code accurately, most attention has been focused on the possibilities within biology, medicine, and bioinformatics. Along with its promise for creating new medicines and gene therapies, the underlying mechanisms of CRISPR have tremendous potential in other fields like designing novel paradigms for hardware security. One benefit CRISPR systems have over other evolutionary development or adaptation systems is their ability to learn and adapt within and across generations. Many evolutionary algorithms rely on large numbers of generations to make gradual changes to arrive at an optimal solution, and while this can be a highly effective method of adaptation and development, CRISPR presents this intergenerational capability along with intragenerational adaptation. The CRISPR system is one mechanism some microorganisms have developed to repel attackers.

CRISPR-Cas systems refer to an array of adaptive, immunity-conferring systems within bacteria and archaea. These systems have been identified in approximately 90% of archaeal genomes and 40% of bacterial genomes [7]. At a basic level, CRISPR is a system with a genetic memory bank of nucleotide sequences the organism has encountered, or its ancestors have encountered. CRISPR systems are composed of three phases: adaptation (also called acquisition), expression, and interference.

### A. ADAPTATION

The adaptation phase saves new genetic material called spacers into the organism's CRISPR locus allowing the organism to "remember" specific genetic sequences it has previously encountered. CRISPR loci are akin to memory banks which store copies of this genetic code. These loci are located either within the organism's genome or in plasmids within the organism. Methods of acquisition vary across different microorganisms; however, this work primarily focuses on the type 1-E CRISPR system of Escherichia coli. CRISPR acquisition mechanisms are typically classified as either naïve or primed. Naïve adaptation is the simpler form of acquisition where any genetic material of sufficient length can be acquired and inserted into the CRISPR locus; however, this method has no filter which can lead to the acquisition of both foreign and host genetic sequences resulting in multiple spacers in the locus which are not capable of immune response [8]. CRISPR-associated proteins Cas1 and Cas2 are responsible for naïve adaptation after they form a Cas1-Cas2 complex which requires 2 Cas1 proteins and 1 Cas2 protein [9]. This complex diffuses through the cell until it encounters foreign DNA from which it selects a PAM (protospacer adjacent motif) and a protospacer. After the spacer is processed, it is placed into the CRISPR locus where it is flanked by two repeat sequences which denote where the spacer begins and ends. On the other hand, primed adaptation requires more Cas proteins and a CRISPR RNA to recognize a partially matching target sequence. In primed acquisition, the Cascade complex carries a crRNA sequence which enables it to recognize complementary or near-complementary sequences of foreign DNA. If the crRNA sequence is near-complementary to some sequence in the foreign DNA and the appropriate PAM is expressed after this sequence in the foreign DNA, the Cascade complex will flag it as a protospacer for incorporation into the locus. Within at least certain *E. coli* strains, both naïve and primed adaptation mechanisms are present [10]. The processes occurring during the adaptation phase are shown in Figure 1(A).

### B. EXPRESSION

During the expression phase, crRNAs are produced and bind with Cascade complexes to form Cascade:crRNA complexes which are capable of interference and primed adaptation [11]. To start, a long pre-crRNA sequence is transcribed from the DNA in the CRISPR locus. This pre-crRNA sequence contains transcriptions of both spacers and repeats from the locus. The pre-crRNA sequence is processed and cleaved into smaller units which contain a full spacer transcribed sequence flanked by partial repeat sequences on both sides. The mature crRNA then binds with a Cascade complex to begin its search for complementary genetic material. The expression processes are displayed in Figure 1(B).

### C. INTERFERENCE

As Cascade:crRNA complexes search the cell via diffusion, they encounter foreign genetic sequences. If the crRNA sequence carried by this complex is complementary or near-complementary to part of the foreign DNA and the foreign DNA has the appropriate PAM, or protospacer adjacent motif, the Cascade complex will bind to the target site. After binding, a Cas3 nuclease is recruited to the target site for degradation of the target DNA. This degradation destroys the foreign DNA by cleaving it; however, in Cas3 deficient systems, the Cascade complex will remain bound to the complementary foreign DNA sequence. While this action does not cleave the foreign DNA, it has a similar effect in that expression of the foreign DNA is prevented limiting or eliminating the threat of this foreign DNA. This type of binding is often used in CRISPR experiments for gene silencing where specific sequences are not removed but instead are blocked from expression. If the foreign DNA has no sequence complementary to or near-complementary to the crRNA, the Cascade:crRNA complex continues its search, leaving behind the foreign DNA. The interference processes are shown in Figure 1(C).

### D. THE ROLE OF RNA AND DNA

Deoxyribonucleic acid, or DNA, and ribonucleic acid, or RNA, are nucleic acids which form the building blocks of life. Along with complex carbohydrates, proteins, and lipids, nucleic acids are one of four chief macromolecules vital for all known lifeforms. DNA is a molecule with a double-helix orientation caused by the coiling of two polynucleotide strands around one another. These polynucleotide strands are composed of nucleotides which contain cytosine (C), guanine (G), adenine (A), or thymine (T). These four chemical bases serve as a code which stores the information in DNA. The bases pair up together to form base pairs where A and T are paired, and G and C are paired [12]. Both polynucleotide chains within DNA code for the same activity. Unlike DNA, RNA is a single chain polynucleotide and is often folded upon itself when found in nature. Various RNA molecules are active within cells, and viruses often have their genetic information encoded as RNA within their genome. RNA shares 3 of the 4 chemical bases with DNA with uracil (U) in place of thymine. In the context of this work, RNA and DNA are related via the transcription process whereby an RNA strand is transcribed from a DNA template (as in pre-crRNA formation) and the reverse transcription process whereby a DNA sequence is generated from an RNA template (as in creation of a new DNA spacer in the locus from an RNA sequence via naïve or primed adaptation.)

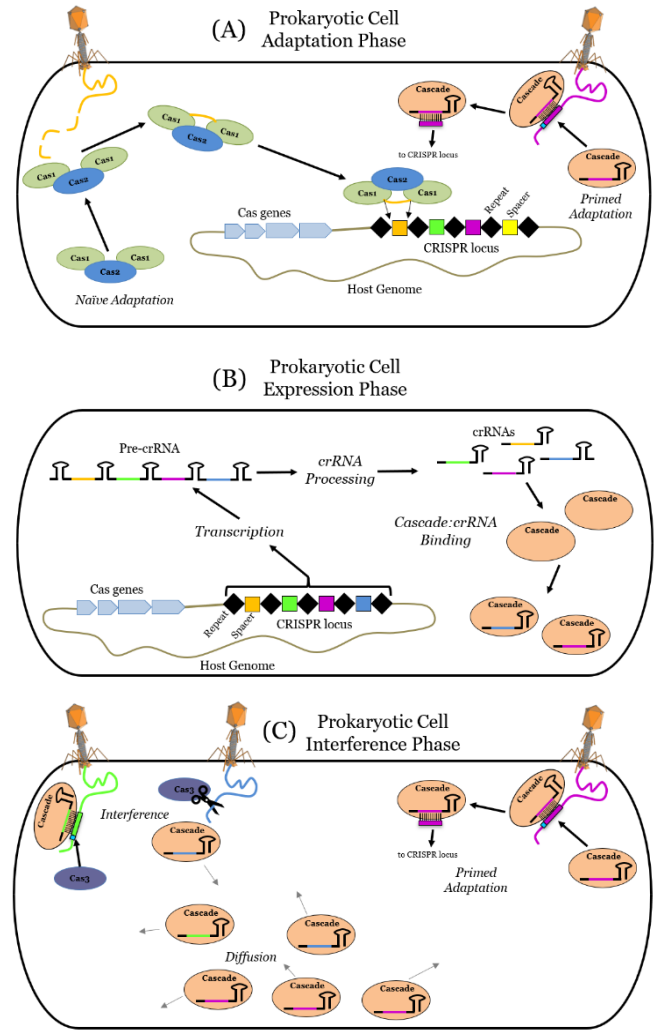

**FIGURE 1.** Phases of the Type 1-E CRISPR-Cas mechanism. (A) shows the adaptation phase including both naïve and primed adaptation. During this phase, spacers are added to the CRISPR locus which provide a memory of previously encountered genetic information. (B) shows the expression phase during which the CRISPR locus is transcribed to crRNAs which guide Cascade complexes. (C) displays the interference phase during which Cascade:crRNA complexes seek out fully or nearly complementary genetic sequences and initiate a response based on this pairing.

### E. PAM SEQUENCE SIGNIFICANCE

The protospacer adjacent motif, or PAM, is a 2-6 nucleotide long sequence which follows the foreign DNA targeted by the Cascade:crRNA complex. These PAM sequences are components of the invading bacteriophage or plasmid which are not parts of the microorganism's CRISPR locus. This is an important detail which ultimately determines how the Cascade:crRNA complex determines what to do with the genetic information it has encountered. If the PAM sequence were to match a nucleotide sequence adjacent to the spacers of the CRISPR locus, the microorganism's Cascade:crRNA complexes would degrade and destroy the CRISPR locus. Instead, the PAM sequences serve to prevent autoimmunity issues like this. Even if the spacer sequence and a subsequence of the foreign genetic material are complementary, without the proper PAM sequences in the foreign genetic material, the Cascade:crRNA complex will not successfully bind to the target, recruit Cas3 for cleaving, or engage in primed adaptation.

## III. FPGA DESIGN

FPGA design can be separated into two distinct product flows. First, the FPGA device itself, or the base array, is designed and manufactured by companies like Xilinx or Altera via now worldwide supply chains. This is the portion of FPGA design which the manufacturer alone is responsible for. The second product flow is the FPGA application design flow which entails the design of the application to run on the FPGA and the configuration of the FPGA to implement this application. Design of this application is carried out by an application developer and is typically accomplished using software tools provided by the vendor. Security is a concern through both of these product design flows.

### A. BASE ARRAY FABRICATION DESIGN FLOW

Design of the FPGA base array follows a standard IC development flow directed by the manufacturer. Using standard design tools and libraries, the base array is designed, fabricated at a foundry, and tested. Following fabrication and testing, the device is normally shipped to a different facility where it is packaged and undergoes final testing. This base array device is then sent to either the final customer or a distributor. Because the base array design follows a standard integrated circuit design and manufacturing process, it is a party to the same issues in supply chain security and trust including reverse-engineering, tampering, and overproduction among others [3]. This serves to illustrate that vulnerabilities exist not only in the application design phase of FPGA design flow but also in the base array design flow.

### B. APPLICATION DESIGN FLOW

The application design flow typically utilizes FPGA vendor tools alongside commercial EDA tools. A visual representation of the application design flow is shown in Figure 3. Preliminary design process steps include the development of project requirements and problem description. Architecture design follows these prerequisites and entails analysis of the project requirements and problem description. This step results in a description of the system's architecture, structure, functions, and interconnects. Next, the architectural design is described in hardware description language (HDL) concurrently with the development of test environments or behavioral models for evaluating the correctness of the system's HDL description. During this step, the application designer might integrate intellectual property and design information from several sources which could include FPGA vendor and third-party libraries, original and repurposed HDL code, and microprocessor software [3]. Following these parallel steps, the correctness of the HDL is verified via comparison to the behavioral model. If there are issues at this stage, the HDL description is revisited to make any necessary adjustments. Once the design passes the behavioral simulation check, the HDL description is synthesized into a netlist which contains the schematic of the digital circuit to be implemented on the FPGA. If there are no issues with synthesis, this netlist can then be mapped to the device's internals which consists of placing and routing which allocates FPGA resources. This is a significant step as it can be viewed as going from general to specific. A generic design in the form of a gate-level netlist may be implemented in any capable FPGA device, but the implementation (placing and routing) configures the information in the netlist for a specific FPGA. After implementation, a timing analysis is performed to ensure the implemented design meets timing requirements. Finally, the bitstream generated during implementation is passed to the FPGA. In general, this process can be broken down into the following levels: behavioral level, register-transfer level (RTL), gate level, and bitstream level. The different steps of the application design flow and the four levels of application design are shown in Figure 2.

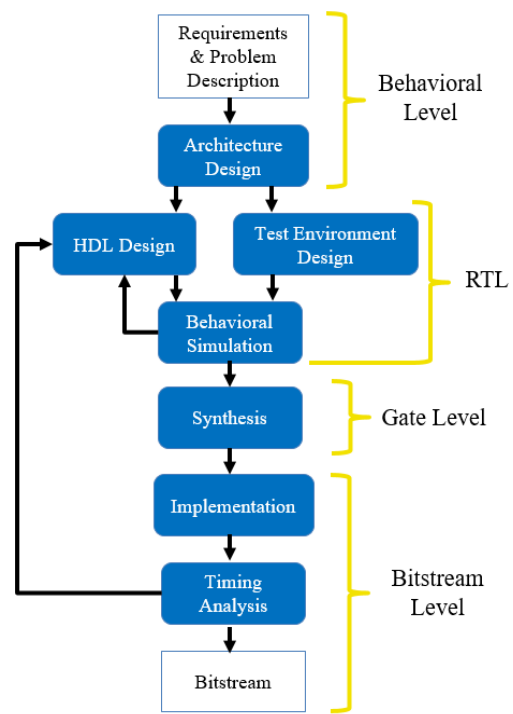

**FIGURE 2. FPGA Design Flow.** The FPGA design flow displayed above depicts the application design process and is broken into four levels: behavioral, register-transfer, gate, and bitstream.

## IV. HARDWARE TROJANS

Hardware Trojans refer to a broad range of malicious alterations made to the circuitry of electronic systems with the goal of yielding some adverse effect. These modifications are typically made without the designer's knowledge. As mentioned before, the supply chain presents many opportunities for Trojan insertion due to the globalization of the semiconductor business. Nowadays, it is rare for hardware design companies to have their own foundry as it is less expensive to outsource this and other stages of the hardware security supply chain to other companies who specialize in one of these areas. While this may bring costs down, it also limits the oversight hardware designers have over the process while increasing the number of players in the supply chain opening the hardware to more attack vectors.

Regarding FPGAs, the global supply chain is slightly different than for some other electronic systems as FPGAs are manufactured by a vendor while hardware designers create the IP which runs on the system. This has been even further outsourced of late with commercially available off-the-shelf (COTS) IP blocks from FPGA vendors. Split manufacturing has been proposed as a potential solution to some of the hardware security issues posed by a globalized supply chain with third-party fabrication; however, split manufacturing alone will not solve challenges to hardware security and is not without drawbacks of its own [13]. Not only does split manufacturing increase the cost and complexity of hardware designs, but it also is not impervious to additional hardware-based attacks including reverse-engineering and Trojan insertion.

Trojans have typically been classified by their trigger and payload. The trigger refers to the way the Trojan is activated which could be internal or external. For example, circumstances which could activate the trigger include an external event like user input or the output of another component or an internal event based on time or the physical condition of the device like its temperature. The Trojan's payload refers to the malicious effects it causes after being activated. Examples of Trojan payload types include device functionality changes, performance downgrades, information leakage, and denial-of-service. In [14], Collins discusses Trojan taxonomies and proposes a new taxonomy for Trojans in FPGA IP to elucidate the threats these Trojans present to hardware designs. He breaks Trojans in FPGA IP into three major segments based on their payload: Trojans that cause malfunction, Trojans that cause side effects, and Trojans that introduce vulnerabilities. Threats to hardware security can occur at various points during the life cycle of a hardware system including during design time, fabrication time, test time, and post-deployment [15]. Bao describes potential threats for each of these phases which can include but are not limited to hardware and software Trojans, IP piracy, IC piracy, overproduction, denial-of-service, recycling, reverse-engineering, brute-force attacks, and side-channel attacks. While many of these threats may be a concern or vulnerability in hardware systems, this work focuses on Trojan-based attacks which can occur at any point during the life cycle of an electronic system like an FPGA. One significant aspect of research into hardware Trojans covers approaches to mitigate their effects which includes designing systems which prevent, detect, or tolerate Trojans. The distinction between the three mitigation methods is significant. Trojan prevention schemes aim to eliminate the possibility of Trojans in hardware via a variety of novel design techniques. Trojan detection schemes enable designers to identify Trojans in their systems during either test-time or run-time; on the other hand, Trojan tolerance strategies enable a hardware system to operate normally despite the presence of one or more Trojans [14]. This work focuses on Trojan detection but also provides some context on Trojan tolerancing methods.

### A. TROJAN INSERTION

Current research suggests that hardware Trojans can be inserted into electronic systems in several different ways. For FPGA devices, Trojan insertion is possible during both the base array design flow and the application design flow. In the base array design and fabrication, Trojans can be inserted at various points in the supply chain including during the design and layout phase via tools used create this design or during fabrication by an attacker with physical access to the device. Trojans in the base array design could be implemented in fewer than 100 gates [16], an extremely small area in modern process nodes. On the other hand, Trojans could be inserted at RTL, gate level, or bitstream level of the application design flow. An exhaustive list of the types of possible Trojans and their insertion methods for both of these design flows is not presented in this work.

### B. TROJAN PREVENTION

Trojan prevention aims to thwart Trojan insertion thus negating the need for detection and subsequent tolerancing or removal. Methods for creating a preventative deterrent include various hardware obfuscation methods like that proposed by Chakraborty and Bhunia in [17]. One major limitation of devoting significant efforts toward Trojan prevention at the expense of detection and tolerancing methods is the incomplete knowledge regarding Trojans and their insertion methods. Comprehensive prevention is a lofty goal which would require extensive and potentially cost-prohibitive examination of each step of FPGA base array fabrication and application design. There is no one catch-all approach to addressing potential issues stemming from hardware Trojans. Alternatively, examining the issue of hardware Trojans from multiple perspectives including prevention, detection, tolerancing, and removal is likely the most effective tactic.

### C. TROJAN DETECTION

Trojan detection relies on the ability to identify potential Trojans in the hardware design; however, this is complicated by the sheer number of possible test vectors for a system like an FPGA. Detection methods can be separated into two categories: destructive and non-destructive. Destructive techniques, like observation under a microscope, are not suitable for Trojans which are hidden in FPGA IP because this technique calls for hardware observation. For FPGA systems, non-destructive techniques include examining code and/or netlists provided by third-party IP vendors, but this requires a significant engineering effort and does not guarantee that a Trojan will be found if it does exist in the IP.

Logical testing is another method of Trojan detection in FPGA IP whereby extensive logical coverage is used. With current testing abilities, it would be infeasible or impossible to test every possible input pattern sufficiently for an FPGA due to exponential scaling as the combinational state-space increases in size. Strategies like MERO [18] aim to make Trojan detection more efficient in a massive test space and maximize device coverage while minimizing testing time. MERO functions via the identification of low likelihood states for every node and the creation of test vectors which intentionally activate those states multiple times. The intuition behind this strategy relies on the assumption that hardware Trojans are likely to be embedded in low probability states to avoid detection during testing, and by probing these low probability states, the detection of Trojans fitting this description would be possible [19]. These logical testing approaches typically utilize a coverage metric to gauge the likelihood of Trojans avoiding detection which may be accomplished via random samples of known Trojans and triggers placed at various points throughout the circuit [18]. Combining the results of these tests enables the development of the aforementioned metric for measuring trigger and detection coverage; however, the test set for this method must be comprehensive so as to provide acceptable coverage [19]. Because it is unlikely that all possible Trojans could be detected during testing, it is desirable to be able to monitor the system during run-time for anomalous behavior which may indicate the presence of a Trojan.

### D. TROJAN TOLERANCING AND REMOVAL

Trojan tolerancing refers to designing with the intent of making the system tolerant of or resilient against Trojans which remain undetected after testing. Two major methods of Trojan tolerancing include run-time monitoring and alternative run-time techniques. Provided the monitoring mechanism is sufficiently comprehensive, run-time monitoring is an effective method for nearly all hardware Trojans [14]. One common technique for run-time monitoring is redundancy with a majority voting scheme like Triple Modular Redundancy (TMR) [20] and related schema. The basis of TMR is the utilization of three redundant, identical versions of the IP which are acquired from different vendors and the assumption that the same Trojan is unlikely to exist in all versions. This method only provides protection from logical errors in the FPGA circuitry caused by a Trojan which would otherwise present an incorrect result in a single FPGA. Information leakage and other malicious activities are still possible with a majority voting redundancy scheme, so various improvements to this mechanism have been proposed. Alanwar et al. propose various modular redundant systems along with Simple Blockage (SB) for data obfuscation [21]. While this may be effective in the mitigation of information leakage Trojans, the implementation of a communication protocol in the IP would preclude the utilization of SB; otherwise, the protocol may not function properly. No current Trojan tolerance systems are capable of resilience against all types of side effect-generating Trojans [14]. Because Trojan tolerancing is normally undertaken by considering only one or a few Trojans, its utility and robustness against a wide array of different Trojans may be more limited than Trojan detection.

## V. A BIOLOGICAL CRISPR-CAS MODEL

While modeling the entirety of a cell or organism like E. coli is certainly an attractive proposition, this capability is currently beyond the reach of modern technology. Whole-cell simulations would be an incredible asset for not only the biology field but also the medical and pharmaceutical fields among others. In 2012, J. Craig Venter Institute researchers reported a basic whole-cell simulation of Mycoplasma genitalium which accounted for all the pathogen's molecular components and their interactions [22]. Despite this step forward, many contended that this simulation only modeled a fraction of the eventual requirements for a realistic simulation of M. genitalium let alone other, more complex microorganisms [23]. This organism has only 525 genes while an organism like E. coli has 4,288 genes and human cells are considerably more complex than even that. The researchers reported that the simulation of a single cell division took approximately 10 hours and generated half of a gigabyte of data [23]. It is almost a certainty that computing technology will continue to advance; however, this anecdote serves to show that at current rates of development, computing technology is a long way from comprehensive whole-cell simulations. Despite these limitations, models and simulations of sub-cell processes will continue to improve offering an unprecedented look into intracellular and intercellular dynamics. The CRISPR mechanism is one such process which has been modeled to better understand how modifications to a cell's genetic code are made.

### A. EXISTING CRISPR-CAS MODELS

Most current CRISPR-Cas models focus on only a subprocess or subprocesses of the CRISPR mechanism as opposed to a complete model which takes all 3 phases into account. The comprehensive biological mechanisms of CRISPR-Cas systems are not currently understood making it impossible to design a full-system implementation which accounts for all dynamics and interactions. Existing simulations like [24] and [25] provide a look into differential equation-based modeling of CRISPR-Cas9 with a focus on the expression and interference stages.

Farasat and Salis [24] designed a CRISPR-Cas9-based model to better characterize the variables which control the system's on-target and off-target activity. Their system-wide biophysical model predicts how variations in "guide RNA sequences, DNA superhelical densities, Cas9 and crRNA expression levels, organisms and growth conditions, and experimental conditions" together regulate cleavage and binding at DNA sites using statistical kinetics and thermodynamics. Expression, Cas9-crRNA complex formation, and DNA target binding are the central components of this model. The model utilizes data from existing experiments to predict off-target binding activity and to better understand the high off-target activity of Cas9. The model results are used to make proposals for limiting off-target activity in future experiments.

Building on the work of [24], Lavington [25] describes inaccuracies in the editing efficiency and off-target effects of biological CRISPR systems. Using similar mechanistic and biophysical parameters, this work investigates targeting efficiency with variations in the guide RNA sequence with a specific focus on R-loop formation which occurs when the Cas9-gRNA complex binds with a DNA site. Using experimental data, the model accounts for position dependence, base pair composition dependence, and consecutive mismatch effects. The plots in Figure 8 below demonstrate the depletion of dCas9 and crRNA molecules through the formation of intermediate complexes. The intermediate complexes eventually settle into the most stable state as demonstrated in the plot of isomerized complex expression. Overall, these plots display the simulation results of a differential equation-based approach to modeling CRISPR-Cas mechanisms. A Gambler's Ruin Markov model was utilized as an analog for R-loop formation while the experimental data was fitted to the model to enable accurate forecasts of bacterial R-loop formations. Although the model simulates CRISPR-Cas9 dynamics during both the expression and interference phases, its primary focus is on a single step of the interference phase.

***B. MODEL CONSIDERATIONS***

The model described in this work is not concerned with the formation and movement of macromolecules and complexes but rather the flow of genetic information through a biological system and its impact on said system. Because of this, it is not intended to be a completely accurate model of all the biological processes which occur before, during, and after the CRISPR process; instead, its purpose is to serve as a basis for examining the mechanisms of the CRISPR process in the context of electronic systems and hardware security. Nevertheless, the model attempts to create a biologically accurate albeit simplified artificial version of the CRISPR-Cas mechanism.

1) PRODUCTION AND DEGRADATION

Within the cell, various proteins, nucleases, and other macromolecules are produced to fulfill myriad functions vital to the cell's normal processes. Some of the key proteins in the type 1E CRISPR-Cas mechanism of *Escherichia coli* include Cas1, Cas2, Cas3, Cse1, Cse2, Cas7, Cas5e, and Cas6e. In naïve adaptation, Cas1 and Cas2 form a complex which recognizes genetic information and prepares it for integration as a spacer in the CRISPR locus. Cse1, Cse2, Cas7, Cas5e, and Cas6e together form the Cascade complex which works cooperatively with Cas 3 to recognize and degrade foreign genetic material identified as complementary to the RNA spacer the Cascade complex carries. The Cascade complex also has the ability to bind with foreign genetic material instead of degrading it thus preventing this foreign genetic material from causing damage to the cell. Additionally, the Cascade complex is a key component of primed adaptation. As with production, degradation of these macromolecules occurs with time and activity. The longer a macromolecule exists and the more active it is, the more quickly it will degrade and require replacement. In a healthy cell, the rates of degradation and production for each macromolecular are similar resulting in a stable population through time. To avoid unnecessary complexity and keeping in mind the end goal of this CRISPR model, macromolecule populations were kept constant and were directly related to the number of spacers added to the CRISPR locus and the number of RNA-guided Cascade complexes searching through foreign genetic material.

2) DIFFUSION

Two types of diffusion are noted in [26]: Brownian motion (self-diffusion) and transport diffusion. Self-diffusion is the stochastic movement of macromolecules in a fluid while transport diffusion is the natural movement of molecules from a region of high chemical potential to a region of low chemical potential. Self-diffusion is caused by macromolecule collision with smaller, rapidly moving molecules like water. In a model which intends to model accurately the specific dynamics of intracellular molecular movements, incorporating both self-diffusion and transport diffusion would be critical. While diffusion does play a central role in the biological CRISPR-Cas system through macromolecular interactions with one another and DNA or RNA, we do not consider diffusion in the simple model of CRISPR-Cas because we are more interested in the flow of and reaction to genetic information. In place of the randomness induced by diffusion, we randomly select which new spacers are introduced into the CRISPR locus. We also assume that, given sufficient populations of a mobile genetic element like a bacteriophage, all Cascade-RNA complexes encounter the mobile genetic element. While this may be a stretch of biological capabilities and a strong assumption about the number of bacteriophage elements present, it simplifies the extension to a hardware security focused CRISPR system.

3) BINDING

Binding processes occur in all 3 phases of the CRISPR-Cas system. In the adaptation phase, Cas1 and Cas1 bind to form a Cas1-Cas2 complex which conducts the naïve adaptation search process. This complex binds with foreign genetic material which it brings to the CRISPR locus. In the expression phase, crRNAs bind with Cascade complexes to form search vectors which seek genetic information complementary to their RNA sequence. In the interference stage, these Cascade-RNA complexes bind with a genetic sequence and either utilize a near complementary sequence for primed adaptation or recruit Cas3 to degrade the genetic sequence. Similar to the diffusion process, binding is assumed and is not directly modeled in the simple CRISPR-Cas model discussed in this work. Intermolecular interactions could be taken into account; however, this would greatly increase the model complexity and add little value to the hardware security goal of the model. Binding affinities between spacer sequences and genetic sequences of an MGE are considered indirectly by PAM comparison which is discussed later.

4) DNA AND RNA

DNA and RNA are the central items of interest in this CRISPR-Cas model. From the beginning of the CRISPR mechanism, DNA and RNA play a vital role in the functionality of the system. In the adaptation phase, genetic sequence fragments are carried by the Cas1-Cas2 complex to the CRISPR locus where the sequence is stored as a DNA spacer sequence flanked by repeat sequences. The CRISPR locus is the library of all genetic sequences the organism has encountered and catalogued via naïve or primed adaptation. The ability to acquire new genetic information and incorporate it into the CRISPR locus forms the backbone of the mechanism's adaptability. The transcription of CRISPR locus spacers from DNA to RNA allows for the binding of RNA sequences to Cascade complexes which protect the cell via interference and primed adaptation. To manage DNA and RNA sequences, the model uses the Biopython package [27]. This package has built in functions for parsing through genetic information stored in various file types as well as transcribing RNA from DNA and back-transcribing DNA from RNA.

The model starts by initializing the CRISPR locus with only a leader sequence and a repeat. The leader sequence, from [7], denotes the beginning of the CRISPR locus and enables consistent and structured additions to the CRISPR locus by the Cas1-Cas2 and Cascade complexes [28]. This leader also provides an easily identifiable starting point for transcription of pre-crRNAs which eventually become the guides for Cascade complexes. Transcription the CRISPR locus into crRNAs was biased slightly toward spacers located close to the leader based on experimental research conducted by Terns and Terns [29]. Selection of these spacers for transcription was accomplished via random selection based on a probability distribution which assigned decreasing probabilities to spacers further from the leader sequence. The transcription relationship between RNA and DNA is shown below in Figure 9.

The repeat sequence denotes where spacer sequences begin and end in the CRISPR locus. The repeat sequence utilized in this model, 'GAGTTCCCCGCGCGAGCGGGGATAAACCG' with a length of 29 nt, is described in research conducted by Luo et al. [30]. This repeat sequence is inserted between each spacer in the locus, and together these repeat sequences assist the cell in determining where the foreign genetic information, or spacers, contained in the locus begins and ends. During transcription of the CRISPR locus, a long pre-crRNA sequence is formed which is then cleaved into individual crRNAs which bind with Cascade complexes. These crRNAs contain a full spacer flanked by portions of a repeat. In the model explained here, the spacer is preceded by 8 nucleotides from the repeat, 'AUAAACCG', and is followed by 21 nucleotides, 'GAGUUCCCCGCGCGAGCGGGG', from the repeat as described in [30]. For a spacer of length 32 nt with 4 nucleotide options, there are 18,446,744,073,709,551,616 (over 18 quintillion) possible permutations. This assumes that all possible permutations are valid – something which is not true in nature.

The crRNA sequences are the targeting sequences which the Cascade complexes carry to search for foreign genetic material which may be a threat to the organism. Once the Cascade:crRNA complex encounters foreign genetic material, a check is performed to determine if the crRNA carried by the Cascade complex is complementary to any part of the foreign genetic material. If a subsequence of the foreign genetic material is complementary to the protospacer contained in the crRNA, the Cascade complex flags the genetic material provided the proper PAM sequence is present in the foreign genetic material. In the in vivo CRISPR mechanism of *E. coli*, this would result in either binding with the Cascade complex preventing expression of the information contained in the genetic sequence or recruitment of the Cas3 nuclease for degradation of the foreign genetic material. If a subsequence is only partially complementary to the protospacer contained in the crRNA, the Cascade complex might flag the genetic material for binding/degradation as described above, initiate primed adaptation and insert this nearly matching sequence into the CRISPR locus, or do nothing and continue its search. As with the first degradation/binding scenario with a fully complementary subsequence, the action of the Cascade complex is ultimately determined by the PAM sequence.

5) PAM SEQUENCES

PAM sequences for this CRISPR-Cas biological model were selected from the intermediate and interfering sets of PAMs described in [31]. Musharova et al. demonstrate three distinct variants of PAMs: stable, interfering, and intermediate. The contents of these PAM sets are shown in Table 1. This work considered 3 nucleotide length sequences for a total of 64 possible permutations. The stable set of PAMs was the largest set with 36 variants, and sequences with these PAMs saw stable populations over the duration of the test. With an active CRISPR system, it can be inferred that the CRISPR mechanisms did not interfere with sequences featuring these PAMs or utilize them for primed adaptation. The interfering set of PAMs contained 17 variants, and the CRISPR mechanism successfully interfered with sequences containing these PAMs. They confirmed this by monitoring the population of these interfering PAM-containing sequences which strongly decreased and remained at very low, stable levels. Lastly, the intermediate set contained the remaining 11 variants which saw their relative ratios decrease over time but at a much slower rate than the interfering set. These population dynamics along with a study of the CRISPR locus confirmed that some sequences containing these PAMs were being interfered with by the CRISPR mechanism while others were added to the CRISPR locus via primed adaptation. This intermediate set is significant due to its ability to allow the E. coli to recognize partially matching sequences and either add them to the locus or promote interference. The capability to add near-complementary spacers to the locus is an example of a fuzzy logic problem in nature and enables the microorganism to adapt to potential mutations in viral DNA which might otherwise allow the viral element to evade the cell's defense mechanisms. This differentiation between self and non-self DNA underpins the success of the CRISPR mechanism and the ability of the microorganism to adapt and survive. In this work, near-complementary refers to the similarity of two genetic sequences. For example, if there are two sequences 20 nucleotides in length that are identical except for one nucleotide, we can say these sequences are 95% similar.

TABLE I
PAM CLASSIFICATION SETS FOR *ESCHERICHIA COLI* [31]

| | |
|---|---|
| Stable | 'ACA', 'ACC', 'ACT', 'AGT', 'CAC', 'CAT', 'CCA', 'CCC', 'CCG', 'CCT', 'CGA', 'CGC', 'CGG', 'CGT', 'CTA', 'CTC', 'CTT', 'GAT', 'GCA', 'GCC', 'GCG', 'GCT', 'GGA', 'GGC', 'GGT', 'GTC', 'GTT', 'TCA', 'TCC', 'TCG', 'TCT', 'TGA', 'TGC', 'TGT', 'TTC', 'TTT' |
| Interfering | 'AAA', 'AAC', 'AAG', 'AAT', 'AGG', 'ATA', 'ATC', 'ATG', 'CAG', 'GAA', 'GAG', 'GGG', 'GTG', 'TAA', 'TAG', 'TGG', 'TTG' |
| Intermediate | 'ACG', 'AGA', 'AGC', 'ATT', 'CAA', 'CTG', 'GAC', 'GTA', 'TAC', 'TAT', 'TTA' |

In the Python model designed alongside this work, the action of the Cascade:crRNA complex takes the PAM sequences into consideration. When this complex encounters a foreign genetic element like the E. coli lambda phage, a search is initiated which compares the crRNA spacer to the DNA of the phage. There are three possibilities of this search which include the identification of a complementary sequence in the phage DNA, the identification of a near-complementary sequence in the phage DNA, or no complementary or near-complementary sequence found. In the first 2 cases, the PAM sequences following the complementary target site are compared to the PAM sets outline in [31]. If the sequence is complementary or near-complementary and the PAM sequence is interfering, the Cascade complex flags the entity for interference. Similarly, if the sequence is complementary and the PAM sequence is intermediate, the Cascade complex flags the entity for interference. If the sequence is near-complementary and the PAM sequence is intermediate, the Cascade complex flags the target sequence for integration into the CRISPR locus. Lastly, regardless of the complementariness of the target sequence, the Cascade complex will do nothing to the entity if the PAM sequence is stable. In conjunction with these PAM and complementary sequence parameters, the action of the Cascade complex is determined. The manner in which the Cascade:crRNA complex targets the sequence of complementary genetic material is shown in Figure 3.

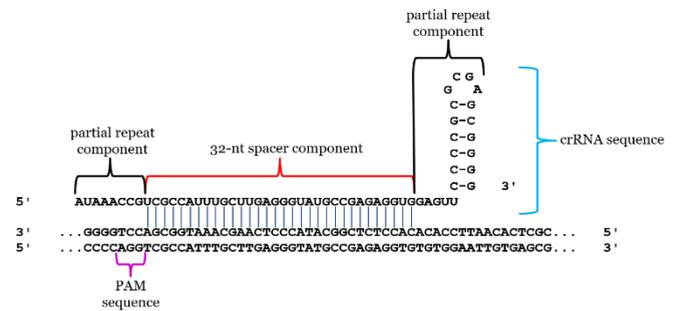

**FIGURE 3. PAM and crRNA Sequences.** The diagram above displays how a crRNA sequence targets complementary foreign genetic material with a specific PAM sequence.

### *C. MODEL DESCRIPTION*

The CRISPR-Cas biological model for simulating the flow of genetic information through the microorganism is structured as follows. First, the CRISPR locus is initialized and the leader sequence and first spacer are inserted. The foreign DNA sequence for a vector like the lambda phage is read in for use later in the network. Next, new spacers are inserted into the CRISPR locus via the naïve adaptation mechanism of the Cas1-Cas2 complex. These randomly selected spacers are acquired from a fragment file which is generated from subsequences of the complete foreign genetic sequence. After each new spacer is inserted, the repeat sequence is appended immediately following the spacer. Once naïve adaptation has occurred, the CRISPR locus is passed to a transcription function which determines which spacer sequences will be transcribed into crRNA to guide the Cascade complexes. Spacers closer to the locus leader sequence are slightly more likely to be transcribed than those further down the locus. Once these spacers are transcribed into crRNA, they are passed to the Cascade action function which compares the crRNA sequences to those of the phage genome to simulate Cascade complexes encountering genetic material from a vector like a bacteriophage. Depending on the PAM sequence and whether the sequence is fully complementary, near-complementary, or not complementary, the Cascade action function counts the instances of interference, primed adaptation, or no action. In the case of no action, the Cascade complex would continue its search for matching or similar genetic sequences in the microorganism.

### D. MODEL RESULTS

This model was tested using DNA from *Escherichia coli* K-12 strain MG1655 [32] and the *E. coli* lambda phage [34]. Sequences 32 nt in length were randomly selected from the self and phage DNA and used to prime the CRISPR locus with spacer sequences via naïve adaptation. These sequences were selected without any regard to the PAMs which followed them in their respective genome. Although sequences of exactly this length may be unlikely and their selection may not be completely accurate from a biological standpoint, this methodology was used to naïvely update the CRISPR locus to enable examination of the Cascade mechanism's efficacy. Multiple tests were conducted using the original lambda phage sequence file, *E. coli* DNA sequence file, phage fragment file, host fragment file, and mutated phage file. Both fragment files contain 1000 32-nt length sequences randomly selected from their original respective files. The mutant phage files contain 50 mutated versions of the original lambda phage file. Each of these mutated variants is the same length as the original phage sequence with a certain percentage of their genetic sequence mutated. For example, a 2% mutation would change 2% of the nucleotides to one of the three other possible DNA nucleotides. These mutated versions were created to simulate and test the interference and primed adaptation mechanisms if the CRISPR system encountered different variants of the same bacteriophage in nature.

The first test involved 2 simulations of the CRISPR-Cas mechanism: one with only phage-derived spacers in the CRISPR locus and another with only host-derived spacers in the locus. The purpose of these simulations, shown below in Figure 4, was to investigate the nature of the naïve adaptation phase of the CRISPR-Cas mechanism. By comparing the interference and no action ratios of the Cascade complex between the two loci, we see on average the mechanism is more likely to interfere with phage DNA than self DNA; however, this alone does not explain CRISPR-Cas-bearing organism's ability to avoid autoimmunity issues. These simulations demonstrate that the naïve acquisition process is more complex than simple random selection of spacers from the environment and is likely linked to specific PAMs. This indicates that the contents of a single PAM set from the 3 sets described in [31] are likely unequal in their ability to promote naïve adaptation.

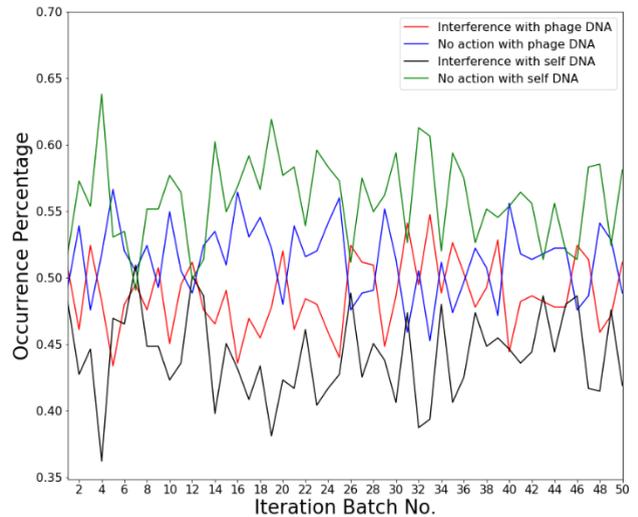

**FIGURE 4.** Interference Rates Between Phage DNA and Self DNA. The figure displays the ratio at which the Cascade complex initiates interference or takes no action for phage DNA and self DNA

In total, interference occurred 44% of the time when the spacers in the locus were solely from the host genome and nearly 49% of the time when the spacers were solely from the lambda phage genome. Correspondingly, no action occurred 56% of the time with host genome spacers and 51% of the time with phage genome spacers. Primed adaptation did not occur for these initial simulations because neither genome was mutated or otherwise modified which resulted in no sequences similar enough to locus spacers which may have enabled primed adaptation. Iteration batches contain twenty individual iterations of the CRISPR-Cas simulation, and the corresponding ratios are the average of the twenty iterations' ratios.

Additional simulations were performed to investigate how primed adaptation might work in the CRISPR-Cas mechanism with different strains or mutations of a bacteriophage. Six separate simulations were performed, and the rates of interference, primed adaptation, and no action were monitored for each. The results of these simulations are displayed in Figure 5. The mutation rate of the lambda phage genome was set at 0.5%, 1%, and 2% in separate simulations. The normal PAM sets were used as described in Table 1 in half of these simulations while modified PAM sets were used to compare the performance of the CRISPR-Cas mechanism under two different PAM conditions. The modified PAM sets involved the transfer of all PAMs except 'CCG' from the Stable set to the Intermediate set. The normal PAM sets reflect those of *E. coli* in [31] while the modified PAM sets reflect a biologically-unlikely scenario in which the only PAM sequence preventing interference is that of the CRISPR locus.

As expected, the interference and primed adaptation rates were much higher with the modified PAM sets when compared to the normal PAM sets. Additionally, higher mutation rates led to higher rates of primed adaptation. This can be explained by the ability of the Cascade:crRNA complex to recognize partially complementary sequences, and if the proper PAM is present, engage in primed adaptation. When there are very few mutations, there is little likelihood of primed adaptation due to the high number of entirely non-complementary sequences and prevalence of entirely complementary sequences. Up to a certain mutation percentage, the primed adaptation rate will continue to increase as the mutation rate increases. Beyond this point, the mutated genetic sequences will be too different from the crRNA spacers carried by the Cascade complexes for primed adaptation to occur.

(A)

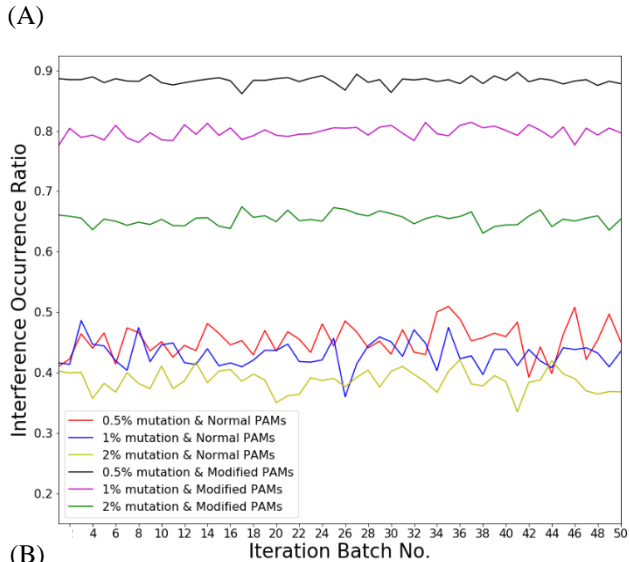

(B)

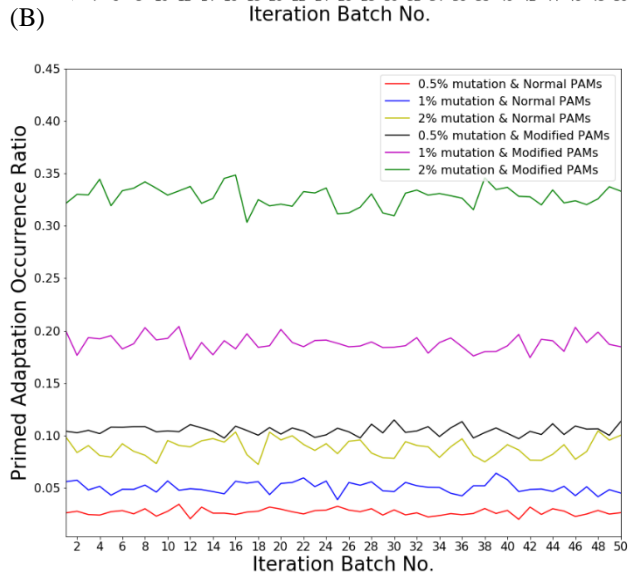

(C)

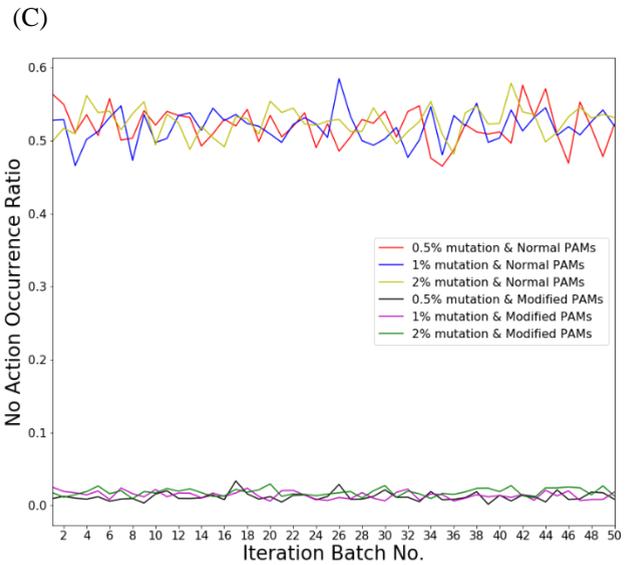

**FIGURE 5.** The Action of the Cascade:crRNA Complex Under Different Conditions. Mutation percentages were set to 0.5%, 1%, and 2% while both normal and modified PAM sets were used. (A) displays the rates of interference with varied mutation rates and PAM sets. (B) displays the rates of primed adaptation with varied mutation rates and PAM sets. (C) displays how frequently no action was taken by the Cascade:crRNA complex with varied mutation rates and PAM sets.

Utilizing the results of the previous tests, the primed spacer acquisition totals and rates can be examined. In Figure 6, the primed acquisition results are displayed. As previous results support, the modified PAM sets promote far higher levels of primed adaptation. Similarly, increasing the mutation rate of the genome to which the locus spacers are compared results in higher primed adaptation spacer counts and therefore a higher acquisition rate.

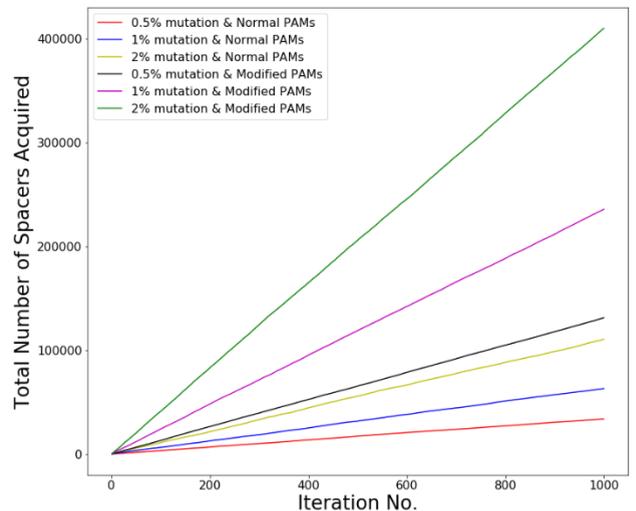

**FIGURE 6.** Primed Spacer Acquisition Counts. The diagram above displays counts of spacers added to the CRISPR locus via primed adaptation for varying mutation percentages and PAM sets.

One significant difference between the simulations presented in this work and the physical CRISPR-Cas system of biological organisms is in the frequency of spacer additions to the CRISPR locus. In general, spacers are added to the locus much less frequently than represented in these simulations [10]; however, this is relevant for its implications in a hardware security method for detection of hardware Trojans. The results obtained from these simulations suggest additional research is required to better explain primed adaptation, specifically the similarity required between crRNA and target for it to occur. A 90% similarity criteria was implemented for these simulations which amounts to a maximum of 3 nucleotide deviations for two 32-nt sequences to be considered similar or near-complementary. The results presented previously also support the notion that naïve adaptation and spacer selection are not well-explained by random acquisition. Although the random acquisition of spacers did indicate that it led to higher levels of interference in foreign genetic information, it is unclear if this would be the case across a range of different bacteriophages, predatory bacteria, or host genomes. Additionally, this random acquisition led to an uncharacteristically high rate of interference with the host's own genome, a significant autoimmunity issue. Three possible explanations for this phenomenon are that the organism very infrequently encounters fragments of its own genome leading to little opportunity to acquire these sequences, or that specific indicators are required to be downstream from the genetic material which could become a spacer, or that the rate of spacer acquisition was substantially lower than that of the simulation as described in [10].

## VI. A CRISPR-CAS-INSPIRED TROJAN DETECTION MECHANISM

Although a comparison was made earlier, it is important to note differences between a microorganism like E. coli and an electronic system like an FPGA. To start, the CRISPR-Cas mechanism of microorganisms does not play a direct role in changes to the organism's own genetic code other than updates of the CRISPR locus; however, in FPGA systems, the bitstream configures the system following HDL synthesis and layout. In the microorganism case, it is almost never desirable for an external (or internal) stimulus to modify the organism's genetic information because even small changes can have large repercussions for the organism's ability to survive. On the other hand, in an FPGA system, it may be desirable for an update to take place which fundamentally alters the configuration or functionality of the system even while the system is running; nevertheless, the ability to adaptably detect Trojans in FPGA systems is a desirable trait which a CRISPR-like mechanism is well-suited to provide.

Despite these differences, there are other similarities between biological systems and hardware systems which have not been discussed to this point. One such similarity concerns the nature of hardware execution and biological functions in a cell. While software execution is generally described in sequential terms, hardware languages are typically grounded in constructs which are carried out in parallel [34]. Similarly, cells carry out many functions in parallel; even different aspects of the CRISPR-Cas system for monitoring foreign genetic material are carried out in this manner. It is then, perhaps, natural to investigate possible applications of microbiology-inspired defense and adaptation mechanisms in the hardware domain. While the system proposed in this chapter is focused more on Trojan detection than tolerance, CRISPR's flexibility allows potential variations on the vanilla detection implementation. One example of a potential variation would have a CRISPR algorithm adapted to work in conjunction with a genetic algorithm designed to tolerate Trojans via evolution like the system described in [14]. This combination of biology-inspired systems could yield a potent, comprehensive system for Trojan detection and tolerance in FPGA devices.

### A. BEHAVIORAL LEVEL REALIZATION

The behavioral level, or algorithmic level, of FPGA design flow refers to the stage at which the design is algorithmic in nature. This is a high-level specification of the design problem where the behavior is uncoupled from things like clock-level timing. At this stage of the FPGA design process, the design is at a higher level of abstraction allowing the designer to look at the scope of the problem with better control over design architecture optimization. The behavioral level is useful for verification by comparing the desired functionality outlined at behavioral level to the implemented functionality on the FPGA device.

The abstract nature of this level of FGPA design limits its candidacy for implementation of a CRISPR-Cas-inspired Trojan detection system. Additionally, any detection method introduced at the behavioral level may be avoided by a Trojan injected by a sophisticated adversary at a later stage of the FPGA design process. Another consideration in determining the efficacy of a potential detection system at the behavioral level is the likelihood of Trojan insertion at the behavioral level. Because the behavioral stage of the process is when the problem is outlined and desired system behavior is specified, it is unlikely that a hardware Trojan would be inserted at this stage. Coupled with the low likelihood of Trojan insertion, the complexity of detecting a Trojan at this level makes the implementation of a detection system impractical at solely behavioral level. This is not to say that the behavioral level has nothing to offer for a hardware Trojan detection mechanism. Rather, utilizing more than one level of the FPGA design process for a Trojan detection system may make the detection system more robust and capable.

### B. RTL REALIZATION

The RTL level of FPGA design specifies circuit characteristics via the operations and data transfer between registers. RTL code is written in HDL and is synthesizable by a synthesis tool. Design at this level uses an explicit clock and has exact timing bounds such that each process is programmed to occur at a specific time. While the likelihood of Trojan insertion at RTL may be slim, a CRISPR-Cas-inspired Trojan detection system implemented at this level could be effective for the RTL synthesis step onward.

Another more feasible implementation is grounded in principles of assertion-based verification and is described at RTL and implemented to function at device run-time. Assertions have a long history of evaluating systems for functional correctness. Assertion-based verification is already frequently used in processor design, so its method and efficacy has already been demonstrated elsewhere. The introduction to Chapter 4 discussed differences between hardware and software, namely that software typically executes sequentially while hardware is built on constructs which execute parallelly. Because of this, assertions in software are evaluated once the program execution reaches the assertion position; however, hardware assertions are evaluated continuously together with the overall system's execution [34]. Property Specification Language (PSL) was developed and standardized for the formal specification of hardware.

In [34], Ngo et al. present a Hardware Property Checker (HPC) written in PSL and synthesized with RTL code to form a dynamic run-time mechanism which monitors IC properties as a Trojan detection system. Their work built on previous research in run-time checking of properties to identify faults; however, they extend the research into the design of an HPC for the purpose of detecting hardware Trojans in a generic IC. The basis of this Trojan detection method is the monitoring of a circuit's vital properties. If a monitored property is anomalous versus the asserted value, the HPC would be able to identify this and do some specified action to notify about the existence of a potential Trojan. An example of part of the property checker is shown in Figure 18. This work proposes a variation on the method presented by Ngo et al. to create an adaptive HPC system for FPGAs. The nature of an HPC is such that it has the capability to detect Trojans in both FPGA IP and hardware so long as they cause anomalies in hardware properties.

HPCs are created with the aid of circuit specifications provided with ICs and from the study of existing hardware Trojans. In the circuit specification portion of HPC design, a CRISPR-Cas system would be possible; however, it may not provide additional benefits over the standard HPC design because circuit specifications for the FPGA itself should not change and have no need for an adaptive monitoring system. The functionality of the circuit implemented on the FPGA could however be monitored via an adaptive monitoring system like CRISPR. Specifications of the configured circuit on the FPGA might change depending on the circuit's purpose and structure. The standard HPC design could detect Trojans in FPGA IP cores without the assistance of an adaptive component through its monitoring of the specifications provided by the manufacturer. Despite this opportunity for implementation of a CRISPR-Cas-based mechanism in circuit property specification, the best option within the HPC structure lies elsewhere. Where a CRISPR-Cas-based variation could provide significant improvement to HPC design is through the monitoring of known Trojan signatures. Obviously, this is limited by the number of Trojans available for study as described earlier. Studying hardware Trojans including their trigger and payload gives an idea and background for identifying their signatures. A traditional HPC would have several of these signatures hard coded into its design which would allow for detection of the Trojan even if the Trojan does not violate the circuit specifications. The drawback of this implementation is that a similar Trojan which is not accounted for in the HPC's instantiation could be missed by the checker. This is the natural location for a CRISPR-Cas-based mechanism due to its ability to recognize similar sequences and store these new sequences for future use. In order to implement this new mechanism, existing Trojans would need to be studied in the context of an FPGA to observe their properties to be able to differentiate them from normal circuit functionality. Successful implementation will rely on the ability to generate sufficiently unique signatures for Trojans which the CRISPR mechanism will be able to pick out. Altogether, the circuit verification component of the HPC would remain the same while the Trojan properties definition component would be supplemented with a CRISPR-Cas-based mechanism to allow real-time updates and similar Trojan recognition. A design flow for a potential CRISPR-Cas-based HPC is shown in Figure 7.

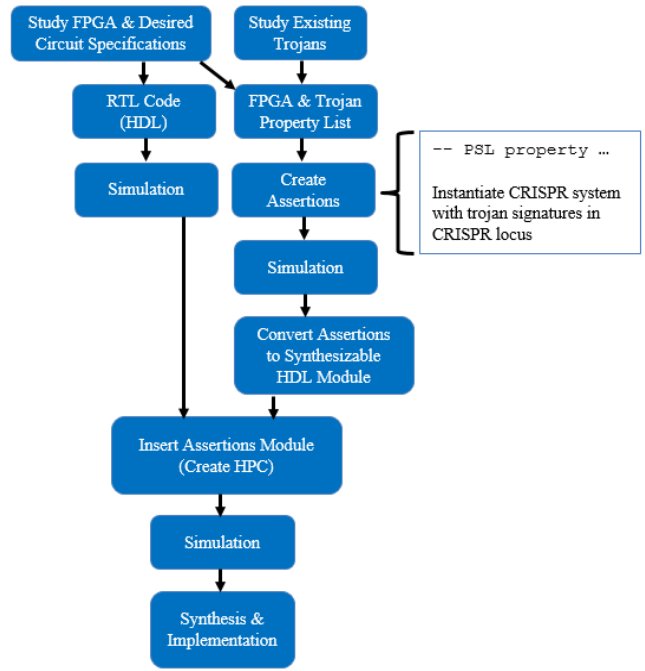

**FIGURE 7.** FPGA Design Flow. The FPGA design flow displayed above depicts the application design process and is broken into four levels: behavioral, register-transfer, gate, and bitstream.

### C. GATE LEVEL REALIZATION

The gate level netlists are generated by a synthesis tool. At this level, the system characteristics are defined through logical links and timing properties, and all signals exist as discrete signals which may only have definite logical values. Possible operations are limited to logic primitives like AND, OR, NOT, etc. Because these netlists are not normally written by designers, they are a more likely vector for Trojan insertion than previous levels of the FPGA design process. Possible methods of Trojan insertion at this level include soft IP cores (both internal and external), but unintentional vulnerabilities introduced in part due to the complexity of the semiconductor manufacturing industry can be equally devastating. Gate level netlists are more difficult for humans to read than HDL code which makes deciphering the exact contents of the netlist more difficult and time intensive.

One proposed solution for detecting Trojans at gate level uses information-flow tracking (IFT) to prove important security properties at gate level (GLIFT) [35]. The GLIFT system is capable of detecting Trojans at RTL and gate level. It does so via formal verification that an information flow maintains confidentiality and integrity-based security properties. Through assignment of bit-level security labels, GLIFT can accurately track information flow through the system preventing untrusted memory locations from accessing things like the cryptographic key. This system represents a step forward from the HPC described previously because GLIFT takes the flow of information into account while formal functional property tools can check that signal A can take value B but do not demonstrate how information flows which is a major consideration in hardware security. Previously, the CRISPR-Cas system of Chapter 3 was shown to model the flow of genetic information through a system. From this perspective, a CRISPR-Cas-based mechanism could fit within the context of the GLIFT system as both are founded on information tracking. GLIFT relies on security theorems created by classification of hardware design signals into varying levels of security. In the vein of GLIFT, a CRISPR mechanism can be adopted to track the information flow in an FPGA. In this implementation, the CRISPR locus would have allowed information flows for comparison with information flows within the circuit. If a fault or Trojan attempts to change an information flow in such a manner that information is compromised, the CRISPR mechanism could detect this change.

Similar to the HDL code analysis proposed previously, a gate level netlist analysis is possible using a system like that of Chapter 3's CRISPR; however, the netlist is a better option than HDL code. Netlists are typically generated by vendor tools provided to function alongside the FPGA. Converting to a netlist from HDL will strip away some of the stylistic differences which can exist in HDL code despite there being no functional difference. If a vendor or FPGA end-user were to incorporate a CRISPR mechanism into netlist analysis, it could potentially identify hardware Trojans or faults before these make it to the on-device implementation. Again, the limitation of this mechanism would be the breadth of hardware Trojans and faults represented in the CRISPR locus and the possibility of high false positive and false negative identifications. A designer could incorporate an analog for PAM sequences into their netlists which would provide a signature for the CRISPR mechanism to recognize as trusted and reduce some of the misidentification, but this PAM analog would take significant consideration to create and would need to be kept confidential. Altogether, the gate level presents some intriguing opportunities for the implementation of a CRISP-Cas-based mechanism for detecting hardware Trojans, but the most straightforward implementation is at bitstream level.

### D. BITSTREAM LEVEL REALIZATION

Bitstream generation is the final stage of the FPGA design process before the FPGA is programmed. The bitstream is generated following placement, routing, and timing analysis. The CRISPR-Cas-based Algorithm for DEtection of FPGA Trojans (CADEFT) analyzes the FPGA bitstream and/or specific hardware properties for known Trojan and fault signatures. CADEFT could be realized in a few different ways: as a part of the synthesis and implementation tools, as an implemented circuit on the FPGA, or as a separate, dedicated chip in an FPGA device. The challenge in detecting Trojans at the bitstream level lies in the opaque nature of FPGA bitstreams. Because descriptions of FPGA device databases are not made available by vendors, the FPGA bitstream formats and entire bitstreams are also unavailable making analysis of bitstreams and hardware security implementations at this level difficult. These realization suggestions would need to be implemented by FPGA manufacturers like Xilinx or utilized on an FPGA with an open source bitstream format. Reverse-engineering of bitstreams for the purpose of creating an open source bitstream generation tool seems to violate vendors' end-user license agreements, and previous tools like JBits which allowed for modification of existing bitstreams are no longer supported by current FPGA families [36]. Researchers have demonstrated that despite incomplete knowledge of bitstream structure, malicious bitstream manipulations by an attacker are not only simpler than presumed but also highly potent and can result in key recovery by the attacker [37].

Realizing CADEFT in the synthesis or implementation tools would allow for detection of Trojans from the gate level and register-transfer level; however, it would not allow for detection of Trojans in the hardware of the FPGA. These Trojans from bitstream level, gate level, or RTL would need to pass through the bitstream to be implemented in the FPGA design; therefore, the bitstream is a natural place to search for and detect them. The CRISPR mechanism has been described sufficiently in previous sections, so an exhaustive explanation of its principles is omitted at this point. This realization requires looking at the bitstream as a sequence of information akin to DNA whose sequences encode instructions for various processes. The generic bitstream structure is shown in Figure 8. When the bitstream is passed to the FPGA device, the configuration does not begin until the synchronization word is encountered which signals the upcoming configuration data. This means that no configuration packet will be processed in the FPGA until the synchronization word is found. Because the FPGA will not begin configuration until this point, we do not need to worry about the header or dummy words. Instead, we can focus on the configuration words as the PAMs and the series of packets as the spacers from a CRISPR perspective. This allows monitoring of the portion of the bitstream which configures the FPGA. When the system encounters a signature in this portion of the bitstream which matches that of a known Trojan or a signature highly similar to a known Trojan, it can send an alert to the user flagging the Trojan's location and its signature from which the trigger and/or payload might be revealed. DNA is more compact from an information storage perspective as the sequences are encoded in a base-4 system while the binary nature of bitstream is a base-2 system. Because of this, it is likely that spacers and PAM sequences for the bitstream would need to be longer, possibly significantly longer, than the 32 nt and 3 nt respective lengths in the biological CRISPR-Cas simulation. Synthesis and implementation tools would be well-equipped to incorporate a CRISPR-like mechanism for bitstream analysis as they are already comprehensive systems, and this could help to mitigate some of the risk inherent in using third-party IP cores.

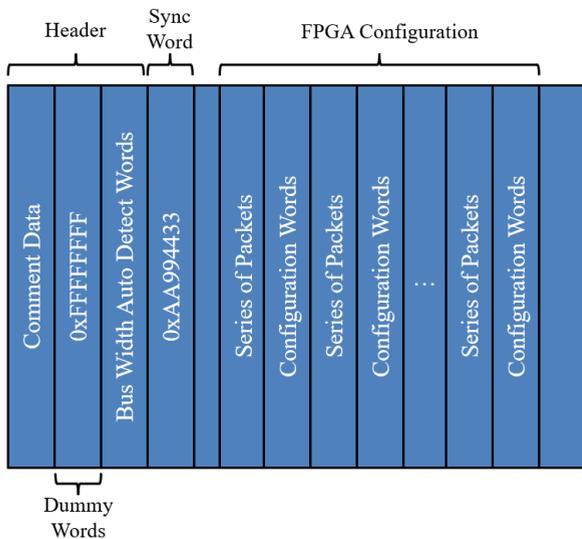

**FIGURE 8.** FPGA Bitstream Configuration. The bitstream format varies by vendor, but all FPGA bitstream configurations contain a header with dummy words, a sync word and the FPGA configuration data.

As an implemented circuit on the FPGA realized alongside the original design, CADEFT could detect Trojans contained in both FPGA bitstream and hardware. This realization would be a step up from the previous suggestion as it has expanded detection capabilities, but it also brings an added cost in the FPGA resources which may already be limited depending on the original design intent for the circuit implemented on the FPGA. This instance of the CRISPR-Cas mechanism would be realized in a similar manner to the hardware property checker described earlier, but this mechanism would examine bitstream level information for Trojans instead of RTL level implementation and synthesis. The bitstream monitoring portion of this realization would be nearly identical to the method discussed for the synthesis/implementation tool, and it would have the added functionality of viewing the real-time signal signatures of the FPGA as sequences for comparing to the signal behavior of known or similar-to-known Trojans. With this setup, the CRISPR-Cas mechanism would have to be implemented in two segments with one monitoring the configuration bitstream and the other focused on the run-time performance of the FPGA. Using benchmarks from trust-hub.org [38], Trojan signatures can be generated via simulation and implementation on real FPGAs. Obviously, simulation signatures would be different from run-time implementation signatures; however, these signatures would not only be useful for realizing the proposed CADEFT system, but they would also be a tremendous asset for future Trojan research.

Realizing CADEFT as an independent chip on the FPGA would certainly be more expensive; however, it would prevent CADEFT from taking up significant FPGA resources while providing a system capable of detecting Trojans introduced via bitstream or in hardware. For this separate realization to be effective, the independent chip would need access to the bitstream and to other hardware properties of the FPGA for monitoring and detection. This is perhaps the most resilient of the three detection methods described in this section as the dedicated chip could not only store information about known Trojans and newly acquired Trojans via a primed adaptation-like mechanism, but it would also maintain a level of isolation from the FPGA which may be desirable to protect the mechanism from Trojans specifically targeting it or from accidental modification of the mechanism during partial FPGA reconfiguration. Should Trojans in hardware be a significant concern, this chip could be produced separately from the rest of the FPGA and inserted at a trusted foundry as is often utilized in split manufacturing techniques. It would also be possible to create this chip in the manner of an FPGA hardware core for easier reconfiguration and updated Trojan signature definitions. This realization would share similarities with a suggestion to create a reconfigurable HPC in [34].

These three different realizations can be implemented in FPGA systems to detect hardware Trojans; however, they would need to be adjusted for different FPGA families as the bitstream composition varies across vendors and FPGA families. This also means that a Trojan which affects one FPGA family may not affect another. In addition to using the signatures of known Trojans in the bitstream, one easy way to at least prime the CRISPR locus of the bitstream-based system would be to give the CRISPR mechanism spacers consisting of bitstream fragments which could physically damage the FPGA. In this manner, the CRISPR mechanism could also monitor potential faults unintentionally introduced into the design which could affect performance or create vulnerabilities.

As mentioned previously, one major bottleneck in Trojan research is the lack of hardware Trojan benchmarks. Trojan examples are currently limited to those designed expressly for academic research like the trust-hub.com benchmarks in Table 2 [39,40]. The lack of hardware Trojans encountered outside the academic realm may seem reassuring, but this does not demonstrate that Trojans are nonexistent in modern supply chains. These benchmarks show that nearly all Trojans designed for research purposes rely on insertion in the design phase with RTL abstraction. While Trojan exploits are certainly possible at other abstraction levels, the RTL is more easily readable by humans than gate or bitstream abstraction levels. Because of the abstraction level funnel which results in the conversion from RTL code to gate level netlists to the bitstream, the bitstream is an optimal place to monitor for Trojans or Trojan activity. Unfortunately, bitstream formats are proprietary, so comparisons of Trojan-infected bitstreams and Trojan-free bitstreams are not available and are thus not presented in this work; therefore, simulation and prototyping of CADEFT are currently outside the scope of this paper. The focal point of this work is to propose a generic CRISPR-Cas-based solution for hardware security in the realm of immunobiology-inspired hardware.

TABLE 2
HARDWARE TROJAN BENCHMARK EXAMPLES FROM TRUST-HUB [39,40]

| Trojan Benchmark | Insertion Phase | Abstraction Level | Location |
|---|---|---|---|
| AES T100 | Design | RTL | Processor |
| AES T1100 | Design | RTL | Processor |
| AES T500 | Design | RTL | Processor |
| AES T900 | Design | RTL | Processor |
| B19-T300 | Design | RTL | Processor |
| BasicRSA-T300 | Design | RTL | Processor |
| memctrl-T100 | Design | RTL | Memory Controller Register File |
| MC8051-T200 | Design | RTL | Processor |
| RS232-T100 | Design | RTL | UART Core |
| wb_conmax-T200 | Design | RTL | I/O |

## VII. FUTURE WORK

This work touched on possible implementations of a CRISPR-Cas-inspired hardware Trojan detection mechanism; however, there are countless other possibilities for utilizing biological systems as inspiration to address issues in electronic systems and create novel improvements. To realize an effective CRISPR-Cas-inspired detection mechanism in hardware, one requisite step is to generate hardware Trojan signatures at the bitstream level. The CRISPR-Cas system functions effectively because spacers fulfill the role of signatures from various sources including phages. Generating Trojan signatures in bitstream format will allow full implementation of the CADEFT mechanism in FPGA devices. While somewhat limited by the availability of hardware Trojans, signature creation would benefit many hardware Trojan research efforts. As an additional step, a CRISPR-Cas-inspired Trojan detection mechanism or even a general Trojan detection mechanism could be made significantly more effective and accurate by examining Trojan signatures at different levels of the FPGA application design flow. By investigating characteristics within signatures at different levels, it may be possible to create clearer distinctions between Trojans and benign sections.

## VII. CONCLUSION

This work discussed several biological mechanisms as well as the issue of Trojans within the hardware security domain. It examined the biological CRISPR-Cas mechanism present in organisms like *Escherichia coli* and how this system is a critical aspect of microorganism defense and adaptation. A biological simulation was presented which attempted to accurately describe and model the various phases of the CRISPR-Cas mechanism including adaptation, expression, and interference. To the writer's knowledge, no simulation or model previously accounted for the flow of genetic information through all phases of the CRISPR mechanism. From creation of the simulation and analysis of the CRISPR-Cas mechanism's functionality, it was determined that a Trojan detection mechanism would be the most apt realization of the CRISPR-Cas system described in this work.

The FPGA design flow was also examined to discern the likelihood of Trojan insertion at various levels and the corresponding fit of a CRISPR-Cas-inspired mechanism for detecting Trojans at these levels. A bitstream-level Trojan detection system was proposed which utilizes the information tracking inherent in the CRISPR-Cas mechanism to monitor FPGA bitstreams for the signatures of known Trojans as well as Trojans which share similarities with known Trojans. The versatility of the proposed CADEFT system is one of its greatest strengths when compared with other Trojan detection methods. The application of subcellular defense and adaptation mechanisms to the field of hardware security represents a novel development in this field.

## ACKNOWLEDGMENT
The authors thank the University of Cincinnati and the Air Force Research Laboratory for providing the support and resources necessary for success. This work was supported by AFRL Award # FA8650-18-C-1191.